\documentclass[aps,prl,twocolumn,groupedadress]{revtex4}
\usepackage{amsmath}
\usepackage{graphicx}

\def\Be+{$^{9}\mbox{Be}^{+}$}
\def\He+{$^{4}\mbox{He}^{+}$}
\def\Mg+{$^{24}\mbox{Mg}^{+}$}
\def\Ca+{$^{40}\mbox{Ca}^{+}$}

\begin{document}

\title{\bf Ellipsoidal Coulomb Crystals in a Linear Radiofrequency Trap}
\author{U.~Fr\"ohlich, B.~Roth, and S.~Schiller}
\affiliation{\it Institut f\"ur Experimentalphysik,
                 Heinrich-Heine-Universit\"at D\"usseldorf, 40225 D\"usseldorf, Germany}

\begin{abstract}
\vspace{0.2cm} A static quadrupole potential breaks the
cylindrical symmetry of the effective potential of a linear rf
trap. For a one-component fluid plasma at low temperature, the
resulting equilibrium charge distribution is predicted to be an
ellipsoid. We have produced laser-cooled Be$^+$ ellipsoidal ion
crystals and found good agreement between their shapes and the
cold fluid prediction. In two-species mixtures, containing Be$^+$
and sympathetically cooled ions of lower mass, a sufficiently
strong static quadrupole potential produces a spatial separation
of the species. \vspace{0.2cm}
\end{abstract}

\maketitle
\begin{figure}[t]
\centering
\includegraphics[width=5.5cm]{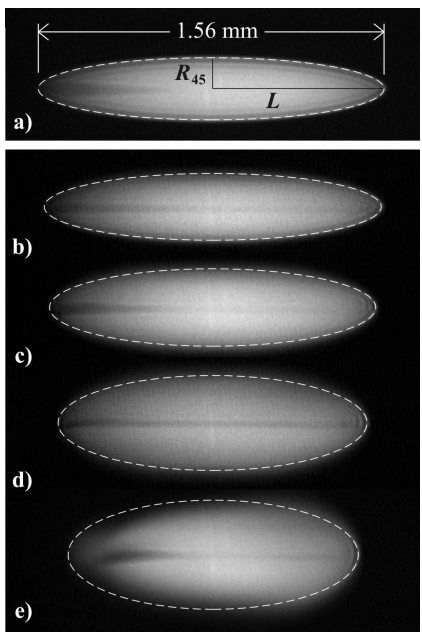}
\caption{CCD images of a large ion crystal containing $\simeq
2.0\times10^3$ \Be+, taken perpendicular to the $z$-axis and at
45$^\circ$ to the $x$- and $y$-axes, for different values of
$V_{DC}$. Dashed lines: fits of ellipses with principal axes
$R_{45}$ and $L$ to the outermost shells. (a): $V_{DC}=0$, the
effective trap potential has cylindrical symmetry. The ion crystal
is a prolate spheroid with radius $R_{x}=R_{y}=R_{45}$ and half
length $L$. The aspect ratio $R_{45}/L = 0.178$. (b)-(e):
$V_{DC}\neq 0$, the cylindrical symmetry is broken and the crystal
is an ellipsoid with principal axes $R_{x}<R_{45}$, $R_{y}>R_{45}$
and $L$. With increasing $V_{DC}$ the crystal expands in the
$y$-direction and compresses in the $x$- and $z$-directions, while
maintaining constant volume. $V_{DC}$ is set to 1.8 (b), 2.8 (c),
3.6 (d) and $4.2\,\mbox{V}$ (e), leading to an aspect ratio of
$R_{45}/L=0.193$ (b), 0.235 (c), 0.309 (d), and 0.413 (e). For a
definition of $R_{x}$, $R_{y}$ and $R_{45}$ see inset of
Fig.\ref{fig_results}. \label{fig_lcrystal}}
\end{figure}
\begin{figure}[t]
\centering
\includegraphics[width=7cm]{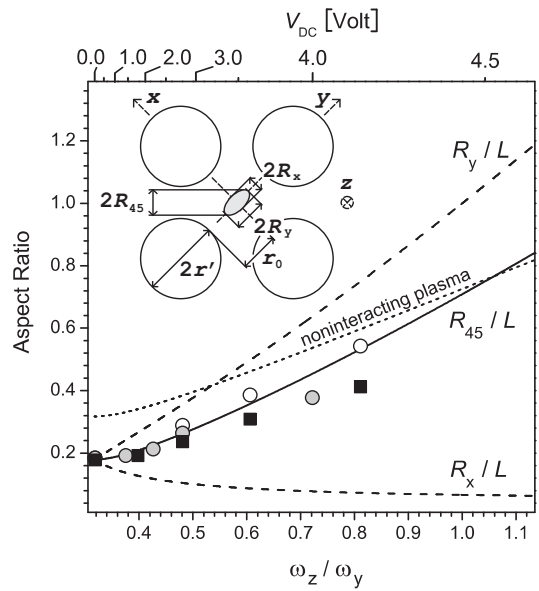}
\caption{Comparison of the aspect ratio $R_{45}/L$ of a partially
crystallized \Be+ plasma and the cold fluid prediction for
zero-temperature ellipsoidal plasmas, as a function of the ratio
$\omega_z/\omega_y$ between the axial and the smallest transverse
trap frequency. Full squares: large crystals from
Fig.\ref{fig_lcrystal}. Open (gray) circles: small (medium-size)
crystals from Fig.\ref{fig_scrystal}. Dotted line: $R_{45}/L$ for
a low-density plasma in the gas phase with negligible particle
interactions, leading to $R_{x}/L=\omega_{z}/ \omega_{x}$,
$R_{y}/L=\omega_{z}/\omega_{y}$. Inset: Cross-section of linear
trap and of ellipsoidal plasma. Observation direction as in
Fig.\ref{fig_lcrystal}. \label{fig_results}}
\end{figure}
\begin{figure}[t]
\centering
\includegraphics[width=7.5cm]{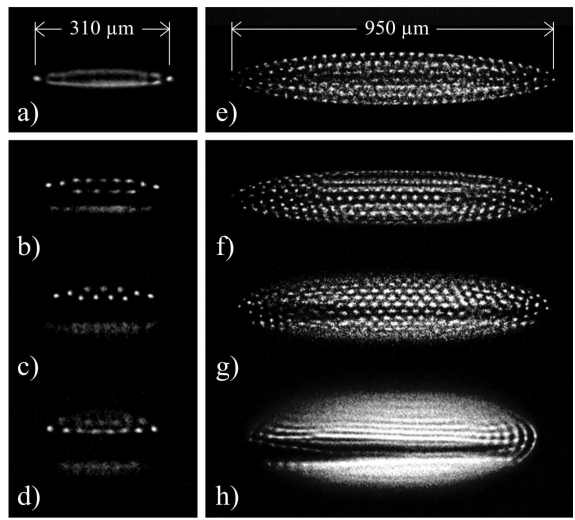}
\caption{Left: small crystal containing $\simeq 20$ \Be+ ions and
a smaller number of sympathetically cooled impurity ions at
different values of the static voltage $V_{DC}$: $0\,\mbox{V}$
(a), $2.8\,\mbox{V}$ (b), $3.6\,\mbox{V}$ (c), and $4.2\,\mbox{V}$
(d). Right: medium-size crystal containing $\simeq 500$ \Be+ ions.
$V_{DC}$ is set to $0\,\mbox{V}$ (a), $1.4\,\mbox{V}$ (b),
$2.8\,\mbox{V}$ (c), and $4.0\,\mbox{V}$ (d). The asymmetric ion
distribution in (b)-(d) and (f)-(h) is due to stray electric
fields. \label{fig_scrystal}}
\end{figure}

One-component plasmas have attracted significant
attention in the past, since they represent simple multi-particle
systems that can be studied under a variety of conditions with a
high degree of experimental control. Detailed theoretical
analysis, both analytical and by molecular dynamics (MD)
simulations, is possible
\cite{Davidson, Dubin1999}.\\
To overcome the Coulomb repulsion, the plasmas are confined in
Penning- or Paul-type traps. Their temperature can be varied over
many orders of magnitude. In particular, they can be efficiently
cooled to the mK range by laser cooling. Strong cooling results in
phase transitions to a crystalline state, whose occurrence is
described by the interaction parameter $\Gamma=Q^2/4\pi\epsilon_0
a k_B T$, the ratio between average nearest-neighbor Coulomb
energy ($a$: average particle spacing, $Q$: particle charge) and
thermal energy. MD simulations on infinite systems have shown that
the plasma becomes fluid, i.e. exhibits spatial correlations for
$\Gamma\geq2$, without going through a discontinuous gas-fluid
phase transition \cite{Dubin1999}. For $\Gamma>170$
\cite{Slattery1980} a phase transition to a crystal occurs.
Coulomb crystals in hyperbolic \cite{Walther1987} and linear Paul
(rf) traps \cite{Raizen1992,Drewesen1998} have become of great
importance in quantum optics, where they can be used to implement
quantum gates or quantum memories \cite{Lukin2000}, and can serve
as systems for precision measurements on atomic or
molecular ions \cite{Berkeland1998, Schiller2003}.\\
The spatial distribution of a trapped one-component gaseous plasma
differs significantly from the fluid and crystalline state. This
is described by the Debye length $\lambda_D=(k_BT\epsilon_0
/nQ^2)^{1/2}$ which is the distance at which interactions between
individual particles overcome collective effects \cite{Dubin1999}.
Here, $n$ is the particle density of the plasma. When $\lambda_D$
is much larger than the spatial extent of the plasma, the Coulomb
interaction is negligible, and the density is a local function  of
the trap potential only. For a harmonic potential, the density has
a Gaussian dependence on the coordinates, with extensions
inversely proportional to the respective trap potential
curvatures. When $\lambda_D$ is comparable to or smaller than the
spatial extent of the plasma (as is the case for the plasmas
presented in this paper), space charge becomes important. The
density is then a local function of both trap potential and space
charge potential. Since the latter depends on the density
distribution over the whole space, a self-consistent density
distribution arises, whose shape exhibits a non-trivial dependence
on the trap potential curvatures. For cylindrical symmetry, such
plasmas are spheroids (ellipses of revolution) and have been
studied both in linear rf and Penning traps
\cite{Hornekaer2001,Brewer1988}. In absence of cylindrical
symmetry the shape has been predicted to be that of an ellipsoid
\cite{Dubin1992}. Ellipsoidal plasmas have already been observed
and studied in Penning traps (see \cite{Huang} and references
therein). In this work we describe the observation and
characterization of cold ellipsoidal plasmas in a linear rf trap.
In particular, we have obtained ellipsoidal crystals containing
two different ion species.\\
The linear rf trap used in this experiment consists of four rods
of radius $r'$ (inset in Fig.\ref{fig_results}), each divided into
three electrically isolated segments. A radio frequency voltage
$V_{RF}\sin\Omega t$ and a static voltage $V_{DC}$ are applied to
the rods in a quadrupolar configuration. Confinement along the
trap axis ($z$-axis) is achieved by raising the two end segments
of each rod by a static voltage $V_{EC}$. When the Mathieu
stability parameter $q=2 Q V_{RF}/m r_{0}^{2}\Omega^{2}\ll 1$, the
independent motion of the trapped ions (mass $m$) can be
adequately described by the motion in an harmonic effective
potential (pseudopotential) $U_{trap}({\bf r})=
m\sum\omega_{i}^{2}x_{i}^{2}/2$ \cite{Dehmelt1967}, and an
additional jitter motion at the radio frequency $\Omega$, the
so-called micromotion. $r_0$ is  the minimum distance from the
electrode surfaces to the trap axis. The axial ($\omega_{z}$) and
transverse ($\omega_{x}$ and $\omega_{y}$) frequencies of the
effective trap potential are given by $\omega_{z}^2 = 2 \kappa Q
V_{EC}/{m}$, $\omega_{x,y}^2= Q^{2} V_{RF}^{2}/({2 m^{2}
\Omega^{2} r_{0}^{4}})-\omega_z^2/2\pm Q V_{DC}/{m r_{0}^{2}}$,
where $\kappa$ is a constant determined by the trap geometry. For
vanishing static voltage $V_{DC}$, $\omega_{x}$ and $\omega_{y}$
are degenerate. With the application of a static voltage
$V_{DC}\neq0$, a static quadrupole potential is added to the
effective trap potential and the cylindrical symmetry is broken.
The transverse trap frequencies $\omega_{x}$ and $\omega_{y}$
increase and decrease, respectively, until $\omega_{y}$ vanishes
for a sufficiently large applied voltage $V_{DC}$, which implies
that the ion motion along the $y$-direction becomes unstable.\\
In the fluid phase, a trapped plasma in thermal equilibrium at a
given temperature $T$ may be regarded as a macroscopic charged
fluid with number density $n_{f}({\bf r})$
\cite{Turner1987,Dubin1999}. This description leads to an
expression for e.g. the equilibrium shape of the trapped plasma. A
particularly simple analytical treatment is possible in the limit
of an ultracold plasma, when $T\rightarrow0$. In this case, the
equilibrium number density $n_{f}({\bf r})$ is determined by the
condition of no net force on any plasma region. However, at very
low temperatures, the plasma does not remain fluid, but
crystallizes with the appearance of shells and a fairly complicate
order. While the fluid description may be expected to become
inaccurate in this case, it has been shown to remain applicable as
far as the shape of the outer boundary is concerned and as long as
the crystals are sufficiently large. For linear rf traps, this
comparison has so far only been performed for the case of cylindrical
symmetry ($V_{DC}=0$) \cite{Hornekaer2001}.\\
In the effective potential approximation, the absence of net force
implies a balance between the electric field due to the space
charge potential $\phi_{f}({\bf r})$ and the trap force due to the
effective potential: $Q\phi_{f}({\bf r})+ U_{trap}({\bf r})=
\mbox{const}$. The number density follows directly from Poisson's
equation, $n_{f}({\bf r})= (\epsilon_{0}/Q^{2}) \Delta
U_{trap}({\bf r})$. For the harmonic effective potential of the
linear rf trap with the above trap frequencies the number density
is explicitly given by $n_{0} = \epsilon_{0} V_{RF}^{2}/{m
\Omega^{2} r_{0}^{4}}$, which is constant within the fluid and
independent of the static voltages $V_{DC}$ and $V_{EC}$.\\
The outer shape of the zero temperature charged fluid with
constant number density is an ellipsoid with principal axes
$R_{x}$, $R_{y}$ and $L$ in the $x$-, $y$- and $z$-direction,
respectively \cite{Dubin1992}. With the additional boundary
condition that the electric space charge potential $\phi_{f}({\bf
r})$ vanishes at infinity, the potential inside the fluid is given
by
\begin{equation}\label{eq_scpot}
\begin{split}
\phi_{f}({\bf r}) = \frac{Q n_{0}}{4 \epsilon_{0}}
                    [ & A_{x}(R_{x}^{2} - x^{2}) + A_{y}(R_{y}^{2} - y^{2}) + \\
                    + & A_{z}(L^{2} - z^{2}) ] \ ,
\end{split}
\end{equation}
\noindent where the dimensionless functions $A_{x}$, $A_{y}$ and
$A_{z}$ depend on the principal axes $R_{x}$, $R_{y}$ and $L$
\cite{Dubin1992}. Force balance implies that the $A_i$ satisfy
$\omega_{i}^{2}=Q^{2} n_{0} A_{i}/2m \epsilon_{0}$. The ratios of
the principal axes $R_{x}/L$ and $R_{y}/L$ can thus be calculated
by solving the set of equations
\begin{equation}\label{eq_axes}
\begin{aligned}
\left( \frac{\omega_{x}}{\omega_{z}} \right)^{2} = &
\frac{A_{x}(R_{x}/L, R_{y}/L)}{A_{z}(R_{x}/L, R_{y}/L)} \\[1ex]
\left( \frac{\omega_{y}}{\omega_{z}} \right)^{2} = &
\frac{A_{y}(R_{x}/L, R_{y}/L)}{A_{z}(R_{x}/L, R_{y}/L)} \ .
\end{aligned}
\end{equation}
\noindent For $V_{DC}=0$, $\omega_{x}$ and $\omega_{y}$ are equal,
and the equilibrium shape of the zero temperature charged fluid is
a spheroid with radius $R:=R_{x}=R_{y}$ and half length $L$. The
two equations (\ref{eq_axes}) then reduce to a single equation for
the aspect ratio $R/L$. Hornek{\ae}r et al.
\cite{Hornekaer2001,Hornekaer2000} have found good agreement
between this theory and experimental aspect ratios of a large
variety of spheroidal \Ca+ ion crystals, obtained by varying the
end-cap voltage $V_{EC}$ and radio frequency amplitude $V_{RF}$.\\
We have tested the predictions of the charged fluid model in a
fully anisotropic effective trap potential using a laser-cooled
\Be+ ensemble. The linear trap properties were
$r_{0}=4.32\,\mbox{mm}$, a radio frequency
$\Omega/2\pi=14.2\,\mbox{MHz}$ with an amplitude
$V_{RF}=380\,\mbox{V}$. This resulted in a small Mathieu stability
parameter of $q=0.055$, implying that the micromotion was
relatively small and that the effective potential description is
appropriate. For axial confinement, $V_{EC}=4.5\,\mbox{V}$ was
applied to the trap end segments, giving rise to $\kappa
=3.0\cdot10^{-3}/\mbox{mm}^{2}$, an axial frequency of
$\omega_{z}=2\pi\cdot85\,\mbox{kHz}$, and a transverse frequency
of $\omega_{r}=2\pi\cdot268\,\mbox{kHz}$. Experimentally, both
values were obtained from a measurement of the transverse
frequency as a function of $V_{EC}$, in absence of $V_{DC}$. This
was done by external excitation of the radial motion of gas phase
\Be+ ions in the harmonic trap potential, which was detected by a
drop of the fluorescence signal from the
laser cooled \Be+ ions \cite{BabaWaki1996}.\\
When the effective trap potential is made anisotropic by
application of $V_{DC}$ the predicted instability limit is
$4.96\,\mbox{V}$. We observe partial particle loss when $V_{DC}$
exceeds $4.2\,\mbox{V}$ and total loss at $V_{DC}=4.9\,\mbox{V}$.
To prevent particle loss, we limited $V_{DC}$ to a maximum of
4.2\,V, corresponding to an increase of the transverse trap
frequency $\omega_{x}/2\pi$ from $268\,\mbox{kHz}$ to
$365\,\mbox{kHz}$ and a decrease of $\omega_{y}/2\pi$ from
$268\,\mbox{kHz}$ to $104\,\mbox{kHz}$. Experimentally, the two
transverse frequencies $\omega_{x}$ and $\omega_{y}$ can be
measured by secular excitation.\\
The trap was loaded with \Be+ ions by evaporating beryllium atoms
from an oven and ionizing them in the trap center by electron
impact. The trapped \Be+ ions, initially forming a hot plasma
cloud of $\approx 1000\,\mbox{K}$, were laser cooled by laser
radiation at $313\,\mbox{nm}$ until they finally underwent a phase
transition to a crystalline state with a temperature of a few mK.
A description of the all-solid-state laser system is given in
\cite{Schnitzler2002}. To image the \Be+ ion crystals, a CCD
camera was placed transverse to the trap axis.
Fig.\ref{fig_lcrystal} shows an ion crystal containing $\simeq
2.0\times10^3$ \Be+, at different values of $V_{DC}$. The estimate
of the ion number in Fig.\ref{fig_lcrystal}~(a) is obtained from
molecular dynamics (MD) simulations in which the observed
structure (especially the number of shells) is reproduced
\cite{Wenz}. For the test of the calibration of the CCD optics
magnification as well as the determination of absolute dimensions
of the ion plasmas, we also use the MD simulations. As a check,
multiplying the volume $4\pi R_{45}^{2} L/3$ of the crystal shown
in Fig.\ref{fig_lcrystal} (a) and the cold fluid model density
$n_{0}$, we obtain the value $\simeq 2.0\times10^3$. This agrees
well with the MD results and implies that the model is applicable
for large crystals.\\
In the outer region of the crystal shown in Fig.\ref{fig_lcrystal}
the ions are arranged in concentric shells, while the core appears
to be fluid. The size of the liquid core appears to increase with
applied static quadrupole potential. A possible explanation for
this behavior is the increased thermal motion of the ions in the
core, which was reproduced by MD simulations and is a subject of
ongoing studies. In contrast to the observation of partially
crystallized large plasmas, we found that under similar laser
cooling parameters small plasmas containing only a few hundred
\Be+ ions always crystallized completely. The left central part of
the ion crystal contains a dark region, which consists of
sympathetically cooled ions originating from the residual gas and
having a mass smaller than that of \Be+, pushed to one side by
radiation pressure \cite{Roth2004}.\\
The ellipsoidal deformation is a reversible process if the maximum
value of $V_{DC}$ is kept within the range stated above: after
turning $V_{DC}$ off, the principal axes $R_{45}$ and $L$ of the
crystal returned to the initial values, indicating that no ions
were lost during deformation. The outer boundaries of the crystals
in Fig.\ref{fig_lcrystal} can be well described by ellipses with
principal axes $R_{45}$ and $L$. Since the CCD camera takes a
projection along an axis at 45$^\circ$ with respect to the $x$-
and $y$-axes, the principal axis $R_{45}$ of each ellipse is
related to the principal axes $R_{x}$ and $R_{y}$ of the
corresponding ellipsoid by
$R_{45}=[(R_{x}^{2}+R_{y}^{2})/2]^{1/2}$. For the last crystal in
the sequence, the boundary shows clear deviations from an ellipse;
we attribute this to the presence of sympathetically cooled
impurities of higher mass than \Be+, located at larger radii as
compared to \Be+. The asymmetry along the trap axis is caused by
cooling light pressure, which is not felt by the sympathetically
cooled ions. In this case, the elliptical fit has been chosen to
match the fragments of the outermost shell, which still exist at
the left and right ends of the crystal.\\
Fig.\ref{fig_results} shows a comparison between the measured
aspect ratio $R_{45}/L$ and the theoretical result from
Eq.(\ref{eq_axes}). The agreement between experiment and theory is
good, considering that the \Be+ ion crystals did exhibit two
phases and were not pure. In addition, a systematic deviation
between theory and experiment is expected at the largest applied
voltages $V_{DC}$ because then the smallest ellipsoid dimension,
$R_x$, becomes comparable to the shell spacing. In this limit, the
cold fluid model continuum description is inaccurate.\\
The cold fluid model also determines the relative change of the
crystal length $2L$ when $V_{DC}$ is changed. For the actual trap
settings $2L$ is expected to decrease to $83\%$ from its initial
value, when changing $V_{DC}$ from 0 to $4.2\,\mbox{V}$. The
length of the crystal in the last image (e) in
Fig.\ref{fig_lcrystal} is $84\%$ of the initial one (a), in good
agreement with the expected compression. Therefore, while at large
$V_{DC}$ the observed transverse crystal shape starts to deviate
from the predictions of the cold fluid model, the axial shape is
still in good agreement, since the axial dimension of the crystal
remains large compared to the shell spacing.\\
While for large Coulomb crystals a spatially averaged description
is a good first approximation, in small crystals one may expect
effects related to the particle structure to show up clearly. As
an example, Fig.\ref{fig_scrystal}~(a)-(d) shows a crystal
containing about 20 \Be+ ions and several sympathetically cooled
low-mass impurities. In an effective trap potential with
cylindrical symmetry, the crystal exhibits a single \Be+ shell,
Fig.\ref{fig_scrystal}~(a). This shell appears smeared out in the
CCD image (exposure time: 2\,s), possibly because of a
rotation-like diffusion of the ions around the trap axis. In the
ellipsoidal crystals, Fig.\ref{fig_scrystal}(b)-(d), this
diffusion is suppressed because it would require overcoming an
energy barrier, and therefore the image shows individual ions. The
most apparent property arising when the static voltage $V_{DC}$ is
increased is the completely dark region containing the
sympathetically cooled particles. Furthermore, some of the
fluorescence spots representing \Be+ ions appear smeared out and
show a reduced intensity. We attribute this to the micromotion
occuring at the locations of these ions. As a test, we shifted the
crystal by means of an additional static voltage applied to one
trap electrode. This caused a rearrangement of the \Be+ relative
to the dark core, but those ions that appeared well defined were
always in the same region.
Although the small crystal (a)-(d) in Fig.\ref{fig_scrystal} does
not exhibit closed boundaries, we could still fit ellipses to
them. In order to compare such small crystals with the cold fluid
model, half of the typical shell spacing of $29\,\mu\mbox{m}$ was
added to $R_{45}$ and $L$ before calculating the ratio $R_{45}/L$.
The results are indicated by open circles in
Fig.\ref{fig_results}. Even for this case there is a good
agreement with the cold fluid model. Finally, in
Fig.\ref{fig_scrystal}~(e)-(h) we also present a medium-sized
crystal containing $\simeq500$ \Be+ ions and, again, additional
low-mass impurities. The corresponding aspect ratio data as well
as an additional data point for $V_{DC}=2.2\,\mbox{V}$, reported
in Fig.\ref{fig_results}, also show good agreement with theory.
However, for large and medium-sized crystals a deviation between
theory and experiment becomes obvious as the values for the static
voltage $V_{DC}$ increase. Our molecular dynamics simulations
show, that the observed deviations can be explained by small ($<
15\%$) admixtures of sympathetically cooled molecular impurities,
in particular, H$_2^{+}$/H$_3^{+}$ ions close to the trap axis
originating from residual gas contaminants, and BeH$^{+}$ ions
located in the outer regions of the crystals formed by chemical reactions.\\
The pronounced asymmetric ion distribution in the small,
Fig.\ref{fig_scrystal}~(a)-(d), as well as in the medium-sized
crystal, Fig.\ref{fig_scrystal}~(e)-(h), is not due to the broken
cylindrical symmetry of the trap potential, as this does not
produce any visible asymmetry in the CCD images. Instead, we
attribute the asymmetric ion distribution to stray potentials.
While it was not possible to compensate for these imperfections by
additional static voltages, it was always possible to reverse the
asymmetry by means of these voltages. Our MD simulations confirm
this interpretation.\\
A direct estimate for the translational temperature of the Be$^+$
is obtained from the spectral line shape of its fluorescence as
the cooling laser is tuned towards resonance and the ion ensemble
crystallizes. Since the temperature of the particles changes
during the frequency scan, we fit a Voigt profile to each point of
the recorded fluorescence curve to determine an upper limit for
the Be$^+$ temperature. For small crystals ($<$1000 particles), we
find an upper limit for the temperature at the end of the scan of
42\,mK. However, the accuracy of this method is limited due to the
experimental resolution. An indirect upper limit is obtained by
comparing the size of the ion spots with MD simulations; here we
find a tighter limit of $<$10 mK for the Be$^+$ temperature. We
deduce, assuming thermal equilibrium, that the temperature of the
sympathetically cooled impurity ions in Fig.\ref{fig_scrystal} is $<$10 mK.\\
In summary, we have studied the static behavior of Coulomb
crystals in a fully anisotropic effective trap potential. We have
found a good agreement with the simple cold fluid plasma model for
small anisotropy. For larger anisotropy, deviations could be
explained by the presence of additional, sympathetically cooled,
ion species. From an experimental point of view, the ability to
reversibly deform a crystal permits to separate lower-mass
sympathetically cooled ions from the laser-cooled ions. This
allows to obtain a clearer picture of the impurity ion ensemble,
without any background or foreground fluorescence from
laser-cooled ions. This will also permit to manipulate the
sympathetically cooled ions in a more direct way. The ability to
generate a variety of ellipsoidal crystals opens up several
directions for further study, e.g. oscillation modes of such
crystals, and in particular the modes of two-species crystals.
These modes could be of importance for identification of the
non-fluorescent species. On the theoretical side, it is of
interest to perform detailed studies of structures using MD
simulations, which are able to take fully into account the
particle nature of the cold plasmas. As an example, our MD
simulations have shown that in strongly squeezed ellipsoids closed
ion rings can occur. These represent a novel form of
low-dimensional artificial structure whose detailed investigation
should be of significant
interest.\\
We thank H.~Wenz for the MD simulations. This work was supported
by the Deutsche Forschungsgemeinschaft and the EU Network
"Ultracold Molecules".\\


\begin{thebibliography}{}

\bibitem{Davidson} R.C.~Davidson, {\it Physics of Nonneutral Plasmas}, (Imperial College Press, 2001).

\bibitem{Dubin1999} D.H.E.~Dubin and T.M.~O'Neil, Rev. Mod. Phys. {\bf 71}, 87 (1999).

\bibitem{Slattery1980} W.L.~Slattery, G.D.~Doolen, and H.E.~DeWitt, Phys. Rev. A {\bf 21}, 2087 (1980).

\bibitem{Walther1987} F.~Diedrich et al.,
                      Phys. Rev. Lett. {\bf 59}, 2931 (1987).

\bibitem{Raizen1992} M.G.~Raizen et al.,
                     J. Mod. Opt. {\bf 39}, 233 (1992).

\bibitem{Drewesen1998} M.~Drewsen el al.,
                       Phys. Rev. Lett. {\bf 81}, 2878 (1998).

\bibitem{Lukin2000} M.D.~Lukin, S.F.~Yelin, and M.~Fleischhauer,
                    Phys. Rev. Lett. {\bf 84}, 4232 (2000).

\bibitem{Berkeland1998} D.J.~Berkeland et al.,
                        Phys. Rev. Lett. {\bf 80}, 2089 (1998).

\bibitem{Schiller2003} S.~Schiller and C.~L\"ammerzahl, Phys. Rev. A {\bf 68}, 053406 (2003),
                       S.~Schiller and V.~I.~Korobov, submitted to Phys. Rev. A.

\bibitem{Hornekaer2001} L.~Hornek{\ae}r et al.,
                        Phys. Rev. Lett. {\bf 86}, 1994 (2001).

\bibitem{Hornekaer2000} L.~Hornek{\ae}r, PhD thesis,
                        Aarhus Univ. (2000).

\bibitem{Brewer1988} L.R.~Brewer et al.,
                     Phys. Rev. A {\bf 38}, 859 (1988).

\bibitem{Dubin1992} D.~H.~E.~Dubin, Phys. Fluids B {\bf 5}, 295 (1992).

\bibitem{Huang} X.-P.~Huang, Phys. Plasmas {\bf 5}, 1656 (1998).

\bibitem{Dehmelt1967} H.~G.~Dehmelt, \newblock Adv. At. Mol. Phys. {\bf 3}, 53 (1967).

\bibitem{Turner1987} L.~Turner, Phys. Fluids {\bf 30}, 3196 (1987).

\bibitem{BabaWaki1996} T.~Baba and I.~Waki, Jpn. J. Appl. Phys. {\bf 35}, L 1134 (1996).

\bibitem{Schnitzler2002} H.~Schnitzler et al., Appl. Optics, {\bf 41}, 7000 (2002).

\bibitem{Wenz} H. Wenz, private communication.

\bibitem{Roth2004} B.~Roth, U.~Fr\"ohlich, and S.~Schiller, submitted.

\end{thebibliography}
\end{document}